\documentclass[a4paper]{article}
\usepackage{amscd,amsmath,amssymb,bbm} 
\usepackage[english]{babel}
\usepackage{enumerate}
\usepackage{pifont}

\numberwithin{equation}{section}

\newcommand{\la}{\lambda}

\newcommand{\g}{\gamma}
\renewcommand{\a}{\alpha}

\renewcommand{\k}{\kappa}
\renewcommand{\o}{\omega}

\newcommand{\s}{\sigma}
\newcommand{\e}{\epsilon}

\newcommand{\Z}{\mathbbm{Z}}
\newcommand{\R}{\mathbbm{R}}
\renewcommand{\P}{\mathbbm{P}}

\newcommand{\C}{\mathbbm{C}}

\newcommand{\N}{\mathbbm{N}}
\renewcommand{\l}{\left(}
\renewcommand{\r}{\right)}
\newcommand{\dd}{{\mathrm d}}

\newcommand{\dz}{\partial_z}

\newcommand{\2}{\frac{1}{2}}                          

\newcommand{\mc}[1]{\mathcal{#1}}
\newcommand{\be}{\begin{equation}}
\newcommand{\ee}{\end{equation}}

\renewcommand{\>}{\rangle}

\renewcommand{\i}{\mathrm i}
\newcommand{\mr}[1]{\mathrm{#1}}

\begin{document}


\begin{titlepage}
\setcounter{page}{0}
\begin{flushright}
      ITP--UH-01/09\\
\end{flushright}
\vskip 2.0cm

\begin{center}
{\LARGE\bf  On a Logarithmic Deformation of the Supersymmetric $bc$-system on 
Curved Manifolds}
\\
\vspace{14mm}

{\Large
Kirsten Vogeler}
\ \ \ and \ \ \
{\Large
Michael Flohr}
\\[5mm]
{ \em
Institut f\"ur Theoretische Physik, Leibniz Universit\"at Hannover \\
Appelstra\ss{}e 2, 30167 Hannover, Germany }
\\[5mm]
email: \texttt{vogeler, flohr @itp.uni-hannover.de}
\\[12mm]
\small \today
\end{center}
\vspace{15mm}


\begin{abstract}
\noindent
E.~Frenkel, A.~Losev and N.~Nekrasov claim that a certain class of theories on compact
K\"ahler manifolds and in particular the ``gauged'' supersymmetric $bc$-system on $\C\P^1$ 
are logarithmic conformal field theories. We discuss that proposition on a 
classical level for the $bc$-system on $\C\P^1$. The outcome of our investigation
conforms to their conjecture. The property of being a {\em logarithmic} CFT thus can be
interpreted as an effect of gravity.

\end{abstract}

\vfill
\end{titlepage}




\section{Introduction}

This paper is the result of our attempt to tackle the following question: can we understand 
the attribute of a conformal field theory on curved target space to be a {\em logarithmic} 
conformal field theory as an aspect of gravity? Below, we will explain the 
background of that question and give an outline.

In 2006, E.~Frenkel, A.~Losev and N.~Nekrasov (``FLN'', if we may) published the 
first part of a 
series of papers (up to now \cite{Frenkel:2006fy, Frenkel:2008vz} and a third part is to 
appear), in which they propose a new perspective on the non-perturbative regime of
certain quantum field theories. In these models, gravitational and topological 
structures are entangled in a non-trivial way, allowing for an analytic solution of the
topological and even of dynamical correlation functions. Thereby the authors extend 
the approach of topological field theories, which did already combine the 
effects of instantons and of curved target spaces, by the dynamical sector.

The type of theory they consider is roughly as follows. The action is a supersymmetric field
theory of type $\mc{N}=2$ in Euclidean dimensions one, two or four, twisted and deformed
to a topological theory with a first order Lagrangian. The target space of the theory is 
some Calabi-Yau or compact K\"ahler manifold. There is an additional topological 
term $S_{top}(\la)$ in the action which triggers 
the contribution of the anti-instanton sector to the correlation functions, depending on
$\la\in\R\ $:
\begin{equation*}
S_\la=S_{\mc{N}=2}+S_{top}(\la)\ .
\end{equation*}
In general correlation functions, the anti-instantons are not damped for $\la=0\ $,
whereas for $0<\la<\infty$ they are damped. Finally, for
$\la=\infty\ $, the instanton sector contributes, while the
anti-instantons are completely damped out. In addition, CPT is broken if $\la\neq 0\ $. 
The topological correlation functions are independent of $\la$ and therefore only the 
dynamical sector is sensitive for the different
``phases'' just described. The authors are mostly interested in the theories with 
$\la=\infty$ because they turn out to be integrable in that phase, even including
dynamical correlation functions.\\
\indent
The domain manifold is assumed to factorize according to $\Sigma_t\times \Sigma_s\ $, where
$\Sigma_t$ is a one dimensional manifold that serves as the domain for the time coordinate.
One can then reduce the theory to a one dimensional super quantum mechanics, a 
Morse-Bott-Novikov theory, by integrating over the space coordinates. Thereby, the authors
obtain a canonical quantized description of the model in which they calculate correlation 
functions as expectation values of (non-)topological observables. What is moreover important
in doing that reduction is that the properties of the one dimensional super quantum 
mechanics should be mirrored in the higher dimensional theory. Therefore, the
higher dimensional field theory can be investigated by means of a more simple model.

If we can say that, in our point of view the piece of work Frenkel, Losev and Nekrasov 
have done is outstanding in the sense that the authors open new 
perspectives on some of the more fundamental questions. One example which is important for 
our investigations is the connection between gravity and topological aspects that can be 
drawn from their work.

The theory they consider is defined on curved target spaces and hence belongs to the field 
of quantum gravity. Usually in quantum gravity one chooses either a path integral approach 
or the method of canonical quantization. FLN's ansatz, however combines both and thus 
serves as a playground for studying the different aspects that can be analysed best by 
each of them, respectively. It turns out, that instanton effects and breaking of 
CPT-invariance are nicely described within the path integral approach, while the question 
of gravity is best treated from the canonical point of view. However, gravity and instantons
are interwoven. They can appear as aspects in different formulations of the 
same structure. Already for the simplest target space and model FLN consider, 
which is Morse theory on $\C\P^1$, it becomes transparent that non-unitarity in the 
non-topological states is such a structure. In the path integral picture, it appears as
an effect of the anti-instantons being absent. In the canonical picture non-unitarity 
emerges in the shape of extra terms in the Hamiltonian as a consequence of $\C\P^1$ being 
compact. This allows for the following question: in what circumstances can
gravity be traded off for instantons and what is the difference in the perspective causing 
that either the one or the other appear?

The question we put is in the same line of substituting gravity by something else that might
be better known. In their theory, Frenkel, Losev and Nekrasov claim that
the models they consider turn out to be logarithmic conformal 
field theories (LCFTs) in the limit of $\la=\infty$ \cite{Frenkel:2006fy, Frenkel:2008vz}. 
The motivation for that conjecture is, that the Hamiltonian of the associated 
Morse-Bott-Novikov theory has Jordan blocks due to the extra terms mentioned above. This is 
exactly the situation in generic\footnote{By generic we mean that 
most studies on LCFTs treat theories that have a non-reducible Hamiltonian, though there 
exist more general situations.} two dimensional LCFTs. These Jordan blocks originate from
the geometry of target space and, to the best of 
our knowlege, this relation between gravity and LCFTs is new, though many connections 
between geometry and conformal field theory have been drawn before.

In order to analyse that connection, we restrict our attention to the supersymmetric
$bc$-system with domain and target space being $\C\P^1\ $. Non-reducibility of the 
Hamiltonian appears in the related one dimensional Morse-Novikov theory.\footnote{An
extra symmetry on target space will be implemented in the action such that we get
rid of the difficulties that come along with the generalizations made by Bott.} 
However, in LCFTs, non-reducibility is a property of the energy momentum tensor or in
mathematical terms of the Virasoro algebra, which is an object in the two dimensional field 
theory. Therefore, in order to analyse the conjecture, we consider it necessary to find 
out, if the derived non-reducibility of the Hamiltonian of the Morse-Novikov theory 
can be obtained by a deformation of the full energy momentum tensor in the two dimensional 
field theory which preserves the Virasoro algebra. This is what we investigated in the following 
sections:

In the first two chapters we will describe the model we are going to investigate and 
derive the operators that deform the energy momentum tensor. We will start with
one of the operators that were already obtained by Frenkel, Losev and Nekrasov in 
\cite{Frenkel:2008vz}. In order to derive the second one, it is sufficient to give some
more detailed arguments, which we will do. Our calculations verify the choices of 
FLN. At the end of chapter \ref{beyond}, we will give a short outline of the procedure 
that will follow and summarize some results.

In chapter \ref{first GCO} we will obtain the deformation operators by means 
of the method of J.~Fjelstad, J.~Fuchs, S.~Hwang, A.~M.~Semikhatov and I.~Y.~Tipunin 
\cite{Fjelstad:2002ei}. The space of 
states of the thus deformed $bc$-system will be described and we also calculate the change 
in the cohomology. A deformation of the supercharge was already calculated by E.~Frenkel 
and A.~Losev in \cite{Frenkel:2005ku} but by a method which usually does not apply to the 
situation under consideration. The result we obtain looks however very similar to theirs.

The last chapter is devoted to a discussion of our results and gives a summary of the
questions that arised during our investigations and that are still open.

\section{The Model under Consideration}\label{model}
Before we start with the subject, we would like to draw the attention of the 
reader to the appendix. Frenkel's, Losev's and  Nekrasov's work connects serveral topics. 
When we wrote this paper we were confronted with the question how self-contained it 
should be. We also had a lot of conventions to choose for the calculations. 
Therefore we decided to make a more detailed
appendix in order to fix the conventions and definitions and to
give very brief introductions to the main
topics ``around'' the subject. A reader who is not an expert in bosonization, the 
chiral de Rham complex or Morse-Bott-Novikov theory might start with the appendix. 
A reader who is well schooled in these things can easily grasp the notations 
and definitions by just taking a glimpse. 

In this section we shortly summarize, how Frenkel, Losev and Nekrasov \cite{Frenkel:2008vz} 
derive the space of states of the model we will discuss here. This is 
important in order to elucidate the connection of the 
underlying Morse-Novikov theory with an associated supersymmetric $bc$-system.
This latter two dimensional CFT will be the subject of our investigations. The
notations and definitions can be found in \ref{chiral bosonization} and \ref{loop space}.

We start with the following action for the situation 
$x: \Sigma\rightarrow X,~\Sigma=\C\P^1=X\ $:
\be\label{gauged a model}
S=\int_{\Sigma} -\i \l p'(\partial_{\bar{z}}+A_{\bar{z}})x -
\pi(\partial_{\bar{z}}+A_{\bar{z}})\psi +c.c.\r\ .
\ee
With non quantized ``gauge field'' $A_{\bar{z}}=\frac{\a}{\bar{z}},~\a
\in(-1,0)\subset\R\ $. 

Due to the $A$-field, the equation of motion is not just the condition for holomorphicity. 
In local coordinates of $\C_0:=X\setminus \{\infty\}$ it reads
\be\label{eom1}
\partial_{\bar{z}}x+\frac{\a}{\bar{z}}x=0\ ,~~~~x(z=0) = 0 = x(z=\infty)\ .
\ee
The boundary conditions are chosen such that the solution $x$ is nonsingular near $z=0$ and 
$z=\infty$, 
where $z\in\Sigma$. They imply that the solutions run into the fixed points 
$\{0,\infty\}$ of some 
$\C^\times$ action with generator $v= x\partial_x + \bar{x}\partial_{\bar{x}}$ on $X$. 

The theory above is transformed to a Morse-Novikov theory on the universal cover of
loop space (c.f.\ section \ref{loop space}). There, one deals with mappings 
$\tilde{\g}: D\times \bar{D}\rightarrow \C_0$ 
that satisfy the Hamiltonian flow equation
\begin{equation}\label{sqm eom1}
\partial_{\bar{z}}\tilde{\g}+\frac{\a}{\bar{z}}\tilde{\g}=0\ ,~~~~
\tilde{\g}(z,\bar{z})\left.\right|_{(0,0)}=0\ ,~z\in D
\end{equation}
and solutions $\tilde{g}: D\times\bar{D}\rightarrow \C_\infty:=X\setminus{\{0\}}$ of
\begin{equation}\label{sqm eom2}
\partial_{\bar{z}}\tilde{g}-\frac{\a}{\bar{z}}\tilde{g}=0\ ,~~~~
\tilde{g}(z,\bar{z})\left.\right|_{(0,0)}=0,~z\in D\ .
\end{equation}
$D\subset \C$ is the disc of radius one. The choice of boundary conditions is now such,
that the solutions $\tilde{\g}$ and $\tilde{g}$ are the flow lines along the descending 
manifold that for $z\rightarrow 0$ run into the fixed points 
$\{0\}\in X$ and $\{\infty\}\in X\ $, respectively.
Analoguous results are obtained for the
superpartners. The solutions of the flow equations above are given by
\begin{equation}
\tilde{\g}(z,\bar{z})=\bar{z}^{-\a}\sum_{n\leq 0}x_n z^{-n}~~~~\mbox{and}~~~~
\tilde{g}(z,\bar{z})=\bar{z}^{\a}\sum_{n< 0}\tilde{x}_n z^{-n}\ ,
\end{equation}
respectively.

Since the local coordinates along the descending manifolds of $\widetilde{LX}$ in the 
related Morse-Novikov theory are identical with the modes of the solutions of the 
respective e.o.m.\ , the local coordinates along 
$\widetilde{LX}_{0,0}$ are $\{x_n\}_{n\leq 0}$ and along $\widetilde{LX}_{\infty,0}$ they 
are the $\{\tilde{x}_n\}_{n<0}\ $.

The operators $x_n$ and $\tilde{x}_n$ are identical with the field modes of the conformal 
field $x(z)\ ,~z\in\Sigma$ in the original theory,
where the tilde denotes the coordinates in the chart $\C_\infty\ $.
Therefore, the {\em naive} space of states,\footnote{What is meant by ``naive'' 
will become transparent in the following section.} associated with the descendent manifolds
of the super quantum mechanics, can be modeled by the chiral de Rham complex of the 
supersymmetric $bc$-system without $A$-field:
\be\label{f zero}
\mc{F}_0=\C[x_n,{p'}_m]_{n\leq 0,m< 0}
\otimes\wedge[\psi_{n},\pi_{m}]_{n\leq 0,m< 0}\cdot|0\rangle_0
\ee
for $\widetilde{LX}_{0,0}$ and for $\widetilde{LX}_{\infty,0}$ it is
\be\label{f one}
\mc{F}^1_\infty=\C[\tilde{x}_n,\tilde{p}_m]_{n< 0,m\leq 0}
\otimes\wedge[\tilde{\psi}_{n},\tilde{\pi}_{m}]_{n< 0,m\leq 0}\cdot|1\rangle_\infty\ .
\ee
The different highest weight vectors originate from the range of the indicees. Indeed,
we obtained two different representations of the Heisenberg and Clifford algebras:
\begin{equation}\label{repr}
\begin{tabular}{ll}
    $x_n|0\rangle_0=\psi_n|0\rangle_0=0\ ,~n>0\ ,$
&$p'_{n}|0\rangle_0=\pi_{n}|0\rangle_0=0\ ,~n\geq 0$\\
    $\tilde{x}_n|1\rangle_\infty=\tilde{\psi}_n|1\rangle_\infty=0\ ,~n>-1\ ,$
&$\tilde{p}'_{n}|1\rangle_\infty=\tilde{\pi}_{n}|1\rangle_\infty=0\ ,~n\geq 1$\\
\end{tabular} 
\end{equation}
Because of the bosons, these representations are not equivalent, c.f.\ appendix 
\ref{chiral bosonization}. 

The naive spaces of states associated with the other sheets, such as 
$\widetilde{LX}_{0,n}\ $, are isomorphic. They are connected by the equivariance 
operator $q$ to the states above: $ q^{n}: ~\mc{F}_0\ni\Psi\mapsto q^n\Psi\in 
q^n\mc{F}_0\ $.

The main observation here is, that the naive space of states of the Morse-Novikov
theory is related to a two dimensional, supersymmetric $bc$-system. That system does
only indirectly know about the $A$-field via the representation spaces as above. 
In the following we will introduce the Grothendieck-Cousin operators that
appear within the particular Morse-Novikov model due to the non-trivial geometry 
of target space. The next step will then be to analyze, how these operators appear as 
(logarithmic) deformations in the energy momentum tensor of the associated $bc$-system. 
This task demands some words of explanation, that we will also shift to the next section.

\section{Beyond Naivity}\label{beyond}
Frenkel's, Losev's and Nekrasov's work \cite{Frenkel:2006fy, Frenkel:2008vz} 
leads to a generalization of the chiral de Rham complex. Their crucial result is that one 
has to include additional operators, Grothendieck-Cousin operators (GCOs), if the space of 
states of 
the Morse-Novikov theory is defined globally on target space. These operators
change the cohomology of the chiral de Rham complex, extend the naive Hilbert spaces and 
deform the Hamiltonian. Roughly, the construction is as follows:

Characterize the descending (or ascending) manifolds. Take their closure, 
for example that of $\widetilde{L\P^1}_{0,n}\ $. Find the descending manifolds that are
contained as a subset of codimension one in any such closure. The naive Hilbert space
$\mc{F}_I\ $, associated with a descending manifold, is then non-canonically extended by
$\mc{F}_{II}$, the naive Hilbert space associated with a descending submanifold of 
codimension one:
\be
0\rightarrow\mc{F}_{II}\rightarrow\mc{F}_{II}\oplus
\mc{F}_I\rightarrow\mc{F}_I\rightarrow 0\ .
\ee 
The GCO is mapping $\mc{F}_I\rightarrow\mc{F}_{II}\ $, while it acts trivially on 
$\mc{F}_{II}$. It appears as an extra term in the naive Hamiltonian, such that
$H = H_{\mr{naive}}+\delta$. Therefore, the Hamiltonian becomes non-diagonalizable on 
certain subspaces.

Another important feature of the GCOs in question is, that they entangle the chiral and 
anti-chiral parts of the model. The reason is that the naive Hilbert spaces are polynomials
in holomorphic and anti-holomorphic coordinates and have to be generalized as
distributions, if they are supposed to be defined globally on target space. The 
Hamiltonian is a Lie derivative in direction of the Morse-Bott-Novikov-potential 
$\nabla f$ and when acting on these polynomial distributions, it relates holomorphic and 
anti-holomorphic coordinates due to the exterior derivative. As a toy model that can be
understood by the relation $\partial_z\bar{z}^{-n-1}\sim\partial_z^n\delta(z)$ on $\C$
for $n\in\N\ $.

To summarize: the nontrivial geometry of the target space causes a deformation of 
the naive space of states. The Hamiltonian is not diagonal on all states, any more, and 
mixes chiral and anti-chiral coordinates. These effects find an expression in additional 
operators, the Grothendieck-Cousin operators, that appear within the Hamiltonian. That
structure is the starting point for some proposals of FLN.

\subsection{On the LCFT Proposal}

Frenkel, Losev and Nekrasov claim \cite{Frenkel:2008vz} that non-reducability of the 
Hamiltonian does also appear in the two dimensional, supersymmetric $bc$-systems on 
Calabi-Yau manifolds. They propose that it is therefore a 
{\em logarithmic} conformal field theory.

In order to analyse that, we must proof that the deformation of the Hamiltonian of the 
Morse-Novikov theory shows up as a deformation of the energy momentum tensor in the 
$bc$-system, transforming it to a {\em logarithmic} conformal field theory. Now we
have to do with two $bc$-systems, the original one (\ref{gauged a model}) and
the associated one at wich we arrived at the end of section \ref{model}. 
The latter differs from the original $bc$-system in that it does only implicitely know about
the $A$-field - it is completely shifted into the spaces of states. This is reflected by 
the fact, that the states in $\C_\infty$ are in the representation $\mc{F}_\infty^1$ and 
not $\mc{F}_\infty^0\ $. If we chose a different value of $\a\ $, we would again get a 
different representation, for instance if $\a\in(0,1)\ $, the r\^ole of (\ref{f zero}) 
and  (\ref{f one}) would be interchanged. To conclude, the zero mode of the energy momentum 
tensor of the associated $bc$-system does not know about the $A$-field, it is just like 
usual the Lie derivative in direction of the generator of loop reparametrizations 
(c.f.\ \ref{chiral bosonization}). 

The Hamiltonian of the Morse-Novikov theory has two parts: it is given by the Lie 
derivative in direction of the sum of the generator of loop reparametrizations as above - 
which hence can be identified with the zero mode of the energy momentum tensor of the 
associated $bc$-system - plus the additional generator that comes from the $\C^\times$ 
action on the target space. As mentioned, that Hamiltonian gains additional GCOs because 
of the exterior derivative $\dd_{LX}\ $, acting on the spaces of states that are globally  
defined on $\widetilde{LX}\ $. The contraction $\iota_{\nabla f}$ does not mix
the spaces of states. Up to some prefactor it acts like $x_n\iota_{\partial_{x_n}}$ for both
symmetries. Therefore, the GCOs do also appear in the associated $bc$-system 
(up to a prefactor that we neglect).

In our analysis we consider that associated $bc$-system on $\C\P^1$ with representation 
spaces $\mc{F}^1_\infty$ and $\mc{F}_0$ on the respective coordinate charts.
We will try to do some deformation 
such that the GCOs enter the zero mode of the energy momentum tensor and investigate, if 
the respective model is an LCFT.

\subsection{On the Cohomology Proposal}

Frenkel, Losev and Nekrasov further propose that the cohomology of the chiral-anti-chiral 
de Rham complex of the associated $bc$-system is deformed by the GCOs. 
In \cite{Frenkel:2005ku} Frenkel and Losev calculated a change of the supercharge, using a 
method outlined in \cite{Zamolodchikov}, from which they say in \cite{Frenkel:2008vz} that
it is exactly the change that stems from the GCOs. This method acts on the level of the two 
dimensional theory and is therefore interesting for our purpose. As the authors did not 
discuss the question of computing the logarithmic correction of the stress tensor, we
are going to do that in the following. However, we choose a different method by wich
we obtain manifest deformation terms in the energy momentum tensor and in the supercharge 
field.

For the model that we just introduced, there exist two Grothendieck-Cousin operators. In 
the next section we will propose a different method in order to deform the $bc$-system
and also let the choice of the GCOs undergo a detailed analysis. Especially for one of
the GCOs that were already obtained in \cite{Frenkel:2008vz} we will fill in the arguments
on the level of the representation spaces. But first, let us specify the algebraic 
structure of the free fermionic $bc$-system in order to identify the Grothendieck-Cousin 
operators.

\subsection{Identification of the Grothendieck-Cousin Operators}
To obtain the Grothendieck-Cousin operators, 
we consider the naive energy momentum tensor of the
associated, free CFT. Mapping it to a supersymmetric $bc$-system, like it is done in 
section \ref{notation}, the energy momentum tensor reads:
\be\label{free tensor}
T(z)=T^+(z)+T^-(z)\ ,
\ee
with zero total central charge. Applying chiral bosonization like in 
\ref{chiral bosonization}, we end up with
\be\label{complete em tensor}
T(z)\mapsto T_{J}(z)=
T_{J^+}(z)+ \l T_{J^-}(z)+T_{\eta\xi}(z)\r \ ,~~~T_{\eta\xi}=:\dz\xi(z)\eta(z):
\ee
and the algebra is that of $\mc{A}^+\otimes\overline{N}(p)\ $. The fields $\eta$ and 
$\xi$ make the auxiliary fermionic $bc$-system that has to be tensored to the bosons in 
order to produce the correct central charge. For the following reason, that system will be 
in the focus of many of our calculations: any effect the Grothendieck-Cousin operator 
does have on the energy momentum tensor must be grounded in the bosons. The mathematical 
argument for that is the relation $M^+(p)\simeq M^+(p'),~\forall~p,p'$ for the fermionic 
sector (c.f.\ \ref{chiral bosonization}).

In \cite{Frenkel:2008vz}, the authors conclude that there must be two Grothendieck-Cousin
operators that are mappings between these spaces: Let
$q$ denote the equivariance operator as explained in \ref{loop space}. Since 
$\overline{\widetilde{LX}}_{\infty,n}\supset \widetilde{LX}_{0,n+1}$ as a
codimension one subset there is a first GCO $\delta_1:~\mc{F}_\infty^1\otimes
\bar{\mc{F}}_\infty^1\hookrightarrow\mc{F}_\times^1\otimes
\bar{\mc{F}}_\times^1 \rightarrow q\l\mc{F}_0\otimes \bar{\mc{F}}_0\r\ $. 
The same applies to $\overline{\widetilde{LX}}_{0,n}\supset\widetilde{LX}_{\infty,n}$ 
and there exists a second GCO $\delta_2:
~\mc{F}_0\otimes\bar{\mc{F}}_0\hookrightarrow \mc{F}_\times\otimes
\bar{\mc{F}}_\times\rightarrow \mc{F}_\infty^1\otimes \bar{\mc{F}}_\infty^1\ $. 
We implemented a coordinate transition from $\C_0\subset X$ to the overlap of both
charges, $\C_\times\subset X\ $, and vice versa, such that for instance 
$\mc{F}_0\hookrightarrow\mc{F}_\times\ $. The GCOs demand that, for they map states 
localized around one pole of $X$ to states that are localized around the other. 
In the following we will specify these two operators.

\subsubsection{The First Grothendieck-Cousin Operator}
The action (\ref{gauged a model}) is formulated in coordinates of $\C_0\subset X$ and
for the moment we take the perspective of an ``observer'' who is ``sitting'' at the fixed 
point $\{0\}$. 

In our ``observatory'' we can bosonize the representation spaces (\ref{repr}) and obtain 
$\mc{F}_0\simeq\mc{A}^+\otimes\overline{N}(0)$ and $\mc{F}_\infty^1\simeq\mc{A}^+
\otimes\overline{N}(1)$, respectively. Since the
fermionic sector is the same for both and since $\xi_0|0\>_{\eta\xi}=|1\>_{\eta\xi}$ while
$\eta_0|1\>_{\eta\xi}=|0\>_{\eta\xi}$, we would naively expect that, after the coordinate 
transition to $\C_\times$, the Grothendieck-Cousin operators are  basically given by 
$\tilde{\eta}_0$, acting on $\mc{F}^1_\times$, and $\tilde{\xi}_0\ $, acting on 
$\mc{F}_\times$. 

We take the first GCO $\tilde{\eta}_0$ for granted, since it is the same as in 
\cite{Frenkel:2008vz}:
\be
\delta_1 := q\oint_{(0,0)}\tilde{\eta}(z)\bar{\tilde{\eta}}(\bar{z})\ .
\ee
Very soon, we will give an additional argument for that choice but first, let us
shortly comment on an obstacle it has: $\mc{F}_\infty^1$ is in the kernel of $\eta_0$ and 
the question appears if that
resists for $\tilde{\eta}_0$ and $\mc{F}_\times^1\ $. Applying the coordinate 
transformation $\C_\infty \mapsto \C_\times$ on the fields according to 
(\ref{trafo rules}), we find
\be
\mc{F}_\infty^1 \hookrightarrow \mc{F}^1_\times=\mc{F}_{0,ferm}\otimes
\mc{F}^1_{\times,bos}\ .
\ee
In terms of the bosonized fields, $\mc{F}^1_{\times,bos}$ is generated by 
$[V^-(+,z)\otimes V_{\eta\xi}^+(+,z)]^{\pm1}\ $,
by $V^-(-,z)\otimes\partial_zV_{\eta\xi}^+(-,z)$ and their derivatives. 
As we explain in \ref{notation}, a transformation $[V^-(+,z)\otimes 
V_{\eta\xi}^+(+,z)]^{-1}\ $ inverts the zero mode. Therefore, the coordinate transformation 
above is equivalent to an inclusion of $\xi_0$ and thus to an extension 
$\overline{N}(1)\hookrightarrow N(1)\ $:
\be
\mc{F}^1_\times \simeq \mc{A}^+\otimes N(1)\ .
\ee 
For convenience, we will omit the tilde on $\eta$ and $\xi$ from now on and hope that the 
context makes clear, in what coordinates we are.

Introducing the operator $\xi_0$ as the second GCO would imply some heavy obstacles. The 
most harmful one is, that it would break conformal symmetry. That can be
seen if we apply some calculation of M.~Krohn and M.~Flohr \cite{Krohn:2002gh}. On the 
mode-level, the authors consider the set of deformations of some supersymmetric 
$bc$-system which preserve the Virasoro algebra. Applying the same calculations to the 
situation under discussion we find, that the conditions of a Virasoro algebra do not allow
for a simultaneous deformation of the Hamiltonian by $\xi_0$ and $\eta_0$. 

Luckily, to choose $\xi_0$ as the second GCO would be wrong and we suggest the following
interpretation why. The reason has to do with the position of our ``observatory'' and with 
non-unitarity of the theory. As we mentioned in 
the introduction, the appearance of the GCOs can not only be understood as an effect of 
gravity but also of the presence of (only !) instantons. In short terms, they mimick the 
instantons, interpolating between different vacua. As CPT invariance is broken, the
anti-instantons are absent, and the instantons flow along the descending manifolds into 
the respective fixedpoints. In our case, sitting at the fixed point $\{0\}\ $, the 
instantons flow from $\{\infty\}$ to $\{0\}\ $, when $z\rightarrow 0\ $. 
The corresponding operator can therefore only be 
$\eta_0\ $. The operator $\xi_0$ represents the outgoing anti-instanton. We obtain the 
corresponding dual, ``anti-GCO'' operator from $\delta_1$ by conjugation and multiplication 
with $q^{-2}\ $, wich is the analogue to \cite[pg.~35, first eqn]{Frenkel:2006fy}. 

Before we derive the second GCO, let us conclude that section with just one remark.
Assumed, that the supersymmetric $bc$-system on target space $\C\P^1$ is an LCFT, the
discussion above links non-unitarity and the question of conformal invariance. It seems, 
that non-unitarity is a necessity for the supersymmetric $bc$-system to be a conformal
theory.

\subsubsection{The Second Grothendieck-Cousin Operator}\label{the second}

The second GCO can only be obtained when changing our ``position'' from $\{0\}$ to an 
``observatory'' at $\{\infty\}\ $. However, we can not do that with a coordinate change
$x\mapsto x^{-1}\ $, alone. The reason is that the action (\ref{gauged a model}) is not
invariant under that mapping, one further has to change the sign of the ``gauge
field''. This is the crucial observation that we owe E.~Frenkel: the second GCO can be 
derived from $\delta_1$ by applying a composition of $x\mapsto x^{-1}$ and a 
rescaling of the $A$-field via $\a\mapsto -\a\ $. In the following we 
will explain the details of that transition up to one open question: unfortunately, we do 
not 
know if, and if yes how, a prefactor $q^{-1}$ is introduced by that composite transition to 
$\{\infty\}\ $, which would pay for the factor of $q$ in $\delta_1\ $. Therefore, we
allow ourselves to add it by hand. The result will be that the second 
GCO is again obtained by $\eta_0\ $, but the space of states
have a different representation in terms of the bosonized theory.

Before we go into details, let us refer the reader to section \ref{in original}, which is
a summary of \cite[pg.~95]{Frenkel:2008vz} in our conventions. Here we obtain the
auxiliary fields $\eta$ and $\xi$ in terms of the original fields. For instance, 
the first GCO reads up to the factor of $q\ $:
\be
\delta_1 \sim - \oint_{(0,0)} \Psi_+(z,\bar{z})\pi(z)\bar{\pi}(\bar{z})\ ,
~~~~
\Psi_\pm(z,\bar{z})=\mr{e}^{\pm\i\l\int^z p'(\o)\dd\o+
\int^{\bar{z}}\bar{p}'(\bar{\o})\dd\bar{\o}\r }\ .
\ee
In the following, we will first apply the transition to $\{\infty\}$ and investigate its
effect on that operator. Afterwards we will discuss the effects on the bosonized
representation spaces.

From the cited appendix we know that the coordinate change $x\mapsto x^{-1}$ 
yields
\be
\delta_1\mapsto - \oint_{(0,0)} \Psi_-(z,\bar{z})\pi(z)\bar{\pi}(\bar{z})\ .
\ee
We further know that on $\overline{N}(p)$ it has the effect $\overline{N}(p)\hookrightarrow
N(p)\ $. Therefore, we concentrate on $\a\mapsto -\a\ $.

The latter causes a change in the charges of the matter fields. This can be understood
as follows. In a generic QFT one introduces an $A$-field by coupling it to a matter current
$j$ such that
\be
\frac{\mr{\delta}S_m}{\mr{\delta}A_\mu}\sim j^\mu\ ,~~~\mbox{where}~~~j^\mu\sim 
  \sum_k~\frac{\partial\mc{L}_m}{\partial(\partial_\mu\phi_k)}~q_k\phi_k\ .
\ee
Here, $S_m$ and $\mc{L}_m$ denote the action and Lagrangian of the matter fields $\phi_k$
with charge $q_k\ $. Usually one absorbs the overall scale factor in the charges and defines
that the left hand side above equals the right hand side. In that respect, a mapping
of $A\mapsto -A$ causes a rescaling of the charges $q_k\mapsto -q_k\ $.

What does that mean for the bosonized system? Let us start with the $bc$-system
as given in \ref{chiral bosonization}. These are the changes:
\begin{equation*}
\begin{tabular}{l|l}
~~~$\a$ & ~~$-\a$\\ \hline
$j^\e(z)=-:b(z)c(z): $&$ \left.j^{\e}\right.'(z) = ~:b(z)c(z):$\\
$j_0|p\>=-\e p|p\>$ &$ \left.j_0\right.'|p\>=\e p|p\> $\\
$\mc{Q} = \e $& $\mc{Q}^\prime  = -\e$
\end{tabular}
\end{equation*}
In particular the OPEs between $b(z)$ and $c(z)$ and the representation 
space $M^\e(p)$ stay the same. Therefore $\a\mapsto -\a$ can only have an effect on the
bosonized system. And the reason for that is that, when we associate it to the already
changed $bc$-system, the new background charge $\mc{Q}^\prime$ enters
the energy momentum tensor and changes the conformal weights. We use the same definitions 
of $J^\e(z)$, of $\phi^\e(z)$ and of the vertex operators. The changes are
\begin{equation*}
\begin{tabular}{l|l}
~~~$\a$ & ~~$-\a$\\ \hline
$\a_0 = -\frac{\e}{2}$ & $\a_0^\prime = \frac{\e}{2}$\\
$\Delta_{T_{J^\e}}\l V^\e(r,z)\r = \frac{\e}{2}~r(r+\e) $&$\Delta_{T_{J^\e}}^\prime\l 
V^\e(r,z)\r =  \frac{\e}{2}~r(r-\e)$\\
$\Delta_{T_{J^\e}}\l \nu_{-\e p}\r =  \frac{\e}{2}~p(p-1) $& 
$\Delta_{T_{J^\e}}^\prime \l \nu_{+\e p}\r =  \frac{\e}{2}~p(p-1)$
\end{tabular}
\end{equation*}
The most important remark is that the identification of states has changed, for now we have 
$\Delta_{T^\e}^\prime(|p\>) = \Delta_{T_{J^\e}}^\prime \l \nu_{\e p}\r\ ,~
J_0\cdot\nu_{\e p} = \left.j_0\right.^\prime |p\>\ $. Hence, the state $|p\>$ gets 
now identified with $\nu_{\e p}\ $, whereas before it was identified with $\nu_{-\e p}\ $.
That will make the change in the representation in the bosonized system. Before we come
to that point, let us derive the correspondence between the fields. As the OPEs stay the
same, we end up with
\begin{equation*}
\begin{tabular}{l|l|l}
$-\a$ &$\e = +$ & $\e = -$ \\ \hline
$b(z)$ & $V^+(+,z)$ & $V^-(+,z)\otimes \partial_z\xi(z)$\\
$c(z)$ & $V^+(-,z)$ & $V^-(-,z)\otimes \eta(z)$
\end{tabular}
\end{equation*}
The auxiliary $\eta\xi$-system is again the same as in \ref{chiral bosonization}. 

In a next step, we can see that the second GCO, that we obtain in the ``observatory''
at $\{\infty\}$ is again basically $\eta_0\ $. Therefore, we consider the changes in the 
representation spaces. While now $M^+(p)\stackrel{\a\mapsto -\a}{\longmapsto} 
M^+(p)\simeq \bigoplus_{l\in\Z} \mc{A}^+_{\2}:= 
\left.\mc{A}^+\right.^\prime\ $, the bosonized bosons make a bigger difference:
\be
M^-(p)\stackrel{\a\mapsto -\a}{\longmapsto} M^-(p)\simeq 
\overline{N}^\prime (p) := \bigoplus_{l\in\Z}\mc{A}^-_{-\2}(-p+l)\otimes
\overline{\mc{A}}^+_{-\2}(l)\ .
\ee
The sign of $p$ is changed on the r.h.s.\ because of the new identification
$|p\>\simeq \nu_{-p}\ $.
Again, $\eta_0 : \mc{A}^+_{-\2}(l)\rightarrow \mc{A}^+_{-\2}(l+1)$ and this time
we get $\eta_0 : N^\prime(p) \rightarrow N^\prime(p+1)\ $. Therefore, $\eta_0$ as
``seen'' from the new ``observatory'' is indeed a mapping 
$\mc{F}_\times \rightarrow \mc{F}^1_0\ $.

If
we now apply the calculation already done in the appendix \ref{in original}, the new GCO 
reads in the original coordinates
\be
\delta_2 := - \oint_{(0,0)} \Psi_-(z,\bar{z})\pi(z)\bar{\pi}(\bar{z})\ .
\ee

\subsection{Summary and Outline of Methods}
We have now obtained both Grothendieck-Cousin operators and have found, that they are
essentially given by $\eta_0\ $, the zero mode of the auxiliary fermionic $bc$-system that
is introduced when bosonizing the theory. Therefore we can concentrate and restrict our
analysis to the situation of the ``observer'' at $\{0\}\ $.

In the following, we will make a logarithmic deformation of the auxiliary $\eta\xi$-system 
in the representation of $\mc{F}_0\otimes\mc{\bar{F}}_0\ $. The method we choose goes back 
to Fjelstad e.al.\ \cite{Fjelstad:2002ei}. 
This will add the GCO $\delta_1$ to the zero mode of the energy momentum 
tensor, which will hence serve as a logarithmic improvement term. All other modes of the
energy momentum tensor will also be affected and the supercharge acquires an 
additional term, as well. 
Furthermore, this sort of deformation will automatically provide us with an 
extension of the algebra 
$\mc{F}_0\otimes\bar{\mc{F}}_0\rightarrow (\mc{F}_0\oplus\left.
\mc{F}^1_{0}\right.^\star)
\otimes(\mc{\bar{F}}_0\oplus\left.\mc{\bar{F}}^1_{0}\right.^\star) \hookrightarrow
\mc{F}_{\eta\bar{\eta}}:=(\mc{F}_0\oplus\mc{F}^1_{\times})\otimes
(\mc{\bar{F}}_0\oplus\mc{\bar{F}}^1_{\times})$\ , where the ``$\star$'' denotes the
kernel of $\xi_0\ $. As we already explained, the space of states 
$\left.\mc{F}^1_0\right.^\star$ is enlarged by applying the coordinate transformation from 
$\C_0$ to $\C_\times\ $. The GCO acts on that latter space.

The result of these examinations will be that the deformation transforms the theory to an 
LCFT with logarithmic partners on each level.

\section{Logarithmic Deformation and the First Grothendieck-Cousin Operator}
\label{first GCO}
We will start with a short summary of the method introduced by \cite{Fjelstad:2002ei}. 
Afterwards, we derive and apply a special version to the auxiliary $\eta\xi$-system in 
order to introduce the first GCO in such a way, that it appears as an improvement term in 
the Hamiltonian.

Fjelstad et al.\ consider a class of deformations of CFTs that lead to logarithmic 
extensions. Basically the idea is to enlarge the conformal algebra by introducing additional
fields, such that the energy momentum tensor gains an improvement term, leading to a 
non-reducible representation.

Let $\mc{C}$ denote some chiral algebra of conformal fields and $\mc{F}$ the corresponding
Fock space with conformally invariant highest weight vector $|0\>_\mc{F}\ $.
We will assume that there exists a fermionic field $E\in\mc{C}$ of conformal
weight one such that $E_0|0\>_\mc{F}=0$ and $E(z)E(\o)=0\ $. The authors deform the
fields $f(z)\in\mc{C}$ by introducing a new field $\delta_E(z)$ and vector 
space $\mc{K}$ such that
\begin{align}\label{defo eq}
  \begin{split}
    \delta_E(z):~\mc{C}\rightarrow \mc{C}\otimes \mr{End}(\mc{K})\ ,~~~
    &f(z) \mapsto~  f_E(z)=:\exp{(-\beta~\delta_E(0))}:f(z)\ ,\\
    \delta_E(z)=1_{\mc{F}}\otimes\delta-\int^z E(\o)\dd\o\otimes 1_{\mc{K}}\ ,~~~
    &\int^z E(\o)\dd\o=E_0~\mr{log}~z-\sum_{n<0}\frac{E_n}{n}z^{-n}-
    \sum_{n>0}\frac{E_n}{n}z^{-n}\ .
  \end{split}
\end{align}
$\delta, \beta \in\mr{End}(\mc{K})$ are Grassman valued. We assume that they
satisfy the condition $[\delta,\beta]=1_{\mc{K}}$ and that we have chosen a vector
$|0\>_{\mc{K}}\in\mc{K}$ such that $\beta|0\>_{\mc{K}}=0$\ . 
The OPE of $\delta_E$ with a field $F(z)=f(z)\otimes \s\ ,~\s\in\mr{End}(\mc{K})$ is given 
by
\be
\delta_E(z)F(\o)=\l -[E,f]_1 \log(\o-z)+
\sum_{n\geq 1}\frac{1}{n}\frac{[E,f]_{n+1}}{(z-\o)^n}\r\otimes \s\ .
\ee
In particular, the
energy momentum tensor gets deformed to
\be
T(z)~\mapsto~ T_E(z)=T(z)+\frac{\beta}{z}E(z)\ .
\ee
That we introduced $\delta_E$ and $\mc{K}$ causes the Virasoro algebra to have a 
non-reducible representation on certain composite fields:
\begin{align}
  \begin{split}
    T_E(z)\Psi_{E,f}(\o)&=\sum_{m\geq 3}\frac{[E,f_E]_{m-1}}{(z-\o)^m}
    +\frac{\Delta_T(f)\Psi_{E,f}+[E,f_E]_1}{(z-\o)^2}+
    \frac{\partial_{\o}\Psi_{E,f}}{z-\o}\ ,\\
    \Psi_{E,f}(z)&:=-:\delta_E(z)f_E(z):\ .
  \end{split}
\end{align}
Since $[\delta,\beta]=1_\mc{K}$\ , the ground state of the extended Fock space is 
degenerate 
by the additional vacuum $E_0^\dagger\delta\cdot|0\>_\mc{F}\otimes|0\>_\mc{K}$\ , where 
$E_0^\dagger$ denotes the momentum conjugate to $E_0$\ . 

Let $|0\>:=|0\>_\mc{F}\otimes |0\>_\mc{K}$ and denote by $\mc{F}'$ the Fock representation
of $\mc{C}$ on that vector. We extended the chiral algebra $\mc{C}$ by introducing the 
additional field $\delta_E(z)\ $. With respect to the representation space, that has the 
effect of introducing an additional state $\delta|0\>\ $. Therefore, the new representation 
space can be identified with $\mc{F}_E:=\mc{F}'\oplus \mc{F}''\ $, where $\mc{F}''$ denotes 
the Fock representation of $\mc{C}$ on $\delta|0\>\ $. The deformed fields act between 
$\mc{F}'$ and $\mc{F}''$\ .

\subsection{Deformation of the $\eta\xi$-System - State of the Art}\label{state of art}
In the following, we denote by $T^{(a)}$ the energy momentum tensor of the auxiliary
$\eta\xi$-system and by $\mc{C}^{(a)}$ the respective chiral algebra of fields. The Fock
representation space is $M^+(0)\simeq\mc{A}^+\ $. The deformation of the 
$\eta\xi$-system as above, where $\eta$ takes the 
r\^{o}le of $E\ $, is already done in \cite{Fjelstad:2002ei}:
\be
  \begin{array}{lll}
T^{(a)}(z)&\mapsto& T^{(a)}_{\eta}(z)=T^{(a)}(z)+\frac{\beta}{z}\eta(z)\ ,\\
\xi(z)&\mapsto&\xi_\eta(z)=\xi(z)+\beta~\mr{log}~z
\end{array}
\ee
and in particular
\be
\left. T^{(a)}_{\eta}\right._0=T^{(a)}_0+\beta~\eta_0\ .
\ee
The new OPE structur is
\begin{align}
  \begin{split}
    \xi_\eta(z)\delta_\eta(\o)&=-\mr{log}(z-\o)\ ,\\
    T^{(a)}_{\eta}(z)\Psi_{\eta,\xi}(\o)&=\frac{0\cdot\Psi_{\eta,\xi}(\o)+ 1}{(z-\o)^2}+
    \frac{\partial_\o\Psi_{\eta,\xi}(\o)}{z-\o}\ ,
  \end{split}
\end{align}
and $\Psi_{\eta,\xi}(z)$ is the logarithmic partner of the identity operator 
$1=1_{\mc{A}^+}\otimes 1_\mc{K}$\ . The energy momentum tensor is therefore degenerate on
the vacuum vector:
\begin{align}
  \begin{split}
    \left. T^{(a)}_{\eta}\right._0\cdot|0\>&=0\cdot|0\>\ ,\\
    \left. T^{(a)}_{\eta}\right._0\cdot\xi_0\delta|0\>&=0\cdot\xi_0\delta|0\>+|0\>\ .
  \end{split}
\end{align}

\subsection{Introducing the Grothendieck-Cousin Operator}
The Grothendieck-Cousin operator is mixing holomorphic and anti-holomorphic
coordinates. We take this as the starting point for a specification of the deformation
just described. The total energy momentun tensor of the $\eta\xi$-system is
\be
T^{(a)}(z,\bar{z})=T^{(a)}(z)+\bar{T}^{(a)}(\bar{z})\ .
\ee
Instead of introducing an abstract vector space $\mc{K}\ $, a sensible way to do a 
deformation mixing the holomorphic and anti-holomorphic parts is, to consider mappings
\begin{align}\label{GC-defo}
  \begin{split}
    \delta_{\eta}:&~\mc{C}^{(a)}\rightarrow\mc{C}^{(a)}\otimes\bar{\mc{C}}^{(a)}\ ,~~~
    \delta_{\eta}(z)=1_{\mc{A}^+}\otimes\bar{\xi}_0-\int^z\eta(\o)\dd\o\otimes 
    1_{\bar{{\mc{A}}}^+}\ ,\\
    \delta_{\bar{\eta}}:&~\bar{\mc{C}}^{(a)}\rightarrow\mc{C}^{(a)}\otimes\bar{\mc{C}}^{(a)}
    \ ,~~~\delta_{\bar{\eta}}(\bar{z})=\xi_0\otimes 1_{\bar{\mc{A}}^+}-
    1_{{\mc{A}^+}}\otimes\int^{\bar{z}}\bar{\eta}(\bar{\o})\dd\bar{\o}
  \end{split}
\end{align}
and field transformations
\be\label{fieldtrafo}
    \delta_{\eta,\bar{\eta}}:~\mc{C}^{(a)}\otimes\bar{\mc{C}}^{(a)} \rightarrow
    \mc{C}^{(a)}\otimes\bar{\mc{C}}^{(a)}\ ,~~~
    f(z,\bar{z}) \mapsto~ f_{\eta,\bar{\eta}}(z,\bar{z})=
    :\mr{e}^{-q\l \delta_\eta(0)~\bar{\eta}_0+
      \eta_0~\delta_{\bar{\eta}}(0)\r }:f(z,\bar{z})\ .
\ee 
This is a specific form of the transformations (\ref{defo eq}) with $\beta=\bar{\eta}_0\ ,~
\delta=\bar{\xi}_0$ and likewise for the transformation on anti-holomorphic fields. We chose
$\delta$ according to the condition $[\delta,\beta]=1_\mc{K}$\ . 
The position of $\eta_0\ $, playing the r\^ole of $\beta\ $, is different from 
(\ref{defo eq}), causing a sign in OPEs.

The mapping above introduces the Grothendieck-Cousin operator
\be
T^{(a)}(z,\bar{z})~\mapsto~ T^{(a)}_{\eta}(z)+\bar{T}^{(a)}_{\bar{\eta}}(\bar{z})
=\l T^{(a)}(z)+\frac{q}{z}~\eta(z)\bar{\eta}_0\r
+\l \bar{T}^{(a)}(\bar{z})+\frac{q}{\bar{z}}~\eta_0\bar{\eta}(\bar{z})\r\ .
\ee
Indeed, consider the original supersymmetric $bc$-system with energy momentum
tensor $T$ as in (\ref{complete em tensor}). We can now apply the deformation
to the auxiliary $\eta\xi$-system, as above. This yields 
$T(z)\mapsto T_\eta(z)=T(z)+\frac{q}{z}\eta(z)\bar{\eta}_0$\ , where
we concentrate on the chiral half. If we now rewrite $\eta$ in terms of the
original fields of the full supersymmetric $bc$-system and integrate over 
$\oint_0 zdz~T_\eta(z)$\ , we end up with exactly the additional term (\ref{GC op}):
\be
T_0~\mapsto~ \left. T_{\eta}\right._0=
T_0 - q\oint_{0,0}~\Psi_+ (z,\bar{z})\pi(z)\bar{\pi}(\bar{z})\ .
\ee

We further have deformations of all other modes and not only of the Hamiltonian. Now again 
in terms of the bosonized fields:
\begin{eqnarray}
T_n ~\mapsto~ \left. T_{\eta}\right._n=T_{n}+q~\eta_n\bar{\eta}_0\ .
\end{eqnarray}


\subsection{Deformation of the Supercharge}
The chiral supercharge gets deformed in the way described above. In terms of the original 
fields we have
\be
\tilde{Q}(z)=Q(z)-\partial_z\psi(z)\ .
\ee
Its OPE with the new field $\delta_\eta(z)$ yields
\be
\delta_\eta(0)\tilde{Q}(z)=-\frac{\Psi_+(z)}{z}\ .
\ee
Therefore, we can calculate the deformation of $\tilde{Q}$ by means of
(\ref{GC-defo}), ending up with:
\be
\tilde{Q}_{\eta}(z)=\tilde{Q}(z)-
\frac{\i q}{z}
\oint_0 \dd\bar{z}~
\Psi_+(z,\bar{z})\bar{\pi}(\bar{z})\ .
\ee
In particular, the zero mode of the supercharge is given by
\be
\left.\tilde{Q}_{\eta}\right._0 = \tilde{Q}_0 + \left.\Psi_+\right._0\bar{\eta}_0\ .
\ee

\subsection{The Structure of the Space of States}
The results of section \ref{state of art} can be carried over to the deformation of
the auxiliary $\eta\xi$-system just considered.
Let $|0\>^{(a)}:=|0\>_{\mc{A}^+}\otimes\bar{|0\>}_{\bar{\mc{A}}^+}$ and again
$\Psi_{\eta,\xi}:=-:\delta_\eta(z)\xi_\eta(z):\ $.  The field $\Psi_{\eta,\xi}$ 
is the logarithmic partner of $1^{(a)}:=1_{\mc{A}^+}\otimes 1_{\bar{\mc{A}}^+}$\ . Hence, 
the energy momentum tensor $T^{(a)}_{\eta}$ has a two dimensional representation on the 
corresponding highest 
weight vector $|1\>^{(a)}:=\xi_0\bar{\xi}_0\cdot|0\>^{(a)}$\ : 
\be
\left. T^{(a)}_{\eta}\right._0\cdot|1\>^{(a)}=-|0\>^{(a)}\ .
\ee
Therefore, the logarithmic partners are modelled on $|1\>^{(a)}$\ . Nevertheless, 
that does not lead to a different representation space, since 
$\mc{A}^+\simeq M^+(1)\simeq M^+(0)$\ , as we explain in appendix \ref{notation}. 

For the supersymmetric $bc$-system we expect a different solution. 
The logarithmic partners are supposed to live in 
$\mc{F}_\times^1\otimes\bar{\mc{F}}_\times^1$\ , while the original
space of states is $\mc{F}:=\mc{F}_0\otimes\bar{\mc{F}}_0$\ . Indeed,
we will find that the zero mode of the
deformed energy momentum tensor $T_\eta(z)$ is a mapping $\left. T_{\eta}\right._0:
~\mc{F}_\times^1\otimes\bar{\mc{F}}_\times^1\rightarrow
\l\mc{F}_\times^1\otimes\bar{\mc{F}}_\times^1\r\oplus q\l\mc{F}_0\otimes\bar{\mc{F}}_0\r$\ .
This affects, however, only a subspace of the full space of states. We will show that
the logarithmic extension leads to the Fock representation
$\mc{F}\rightarrow\mc{F}_{\eta\bar{\eta}}=\l\mc{F}_0\oplus\mc{F}_\times^1\r\otimes
\l\bar{\mc{F}}_0\oplus\bar{\mc{F}}_\times^1\r\ $, where the fields, and in 
particular the energy momentum tensor, are mappings between the summands. 

\subsubsection{The Space of States for the Supersymmetric $bc$-System}\label{space of 
states}
The extension of the space of states of the full supersymmetric $bc$-system
is a bit more complicated than for just the $\eta\xi$-system. The reason 
is that the algebra of the auxiliary fermionic fields does not factorize. 

For convenience, we define $N(p,p'):=N(p)\otimes\bar{N}(p')\ $.\footnote{The overline 
denotes the kernel of $\eta_0$\ , 
while the bar denotes complex conjugation.} 
If one of its factors, say the first, is $\overline{N}(p)$\ , we will write 
$N(\overline{p},p')$ and if it is $N^\star(p)\ $, we will denote it by $N(p^\star,p')\ $.
If both factors are either in the kernel of $\eta_0$ or of $\xi_0\ $, we mark the whole 
thing by $\overline{N}(p,p')$ or $N^\star(p,p')\ $, respectively.

Logarithmic deformation is applied to the bosonic part of the supersymmetric $bc$-system,
which is in the representation $\overline{N}(0,0)$\ . 
A first peculiarity is that
introducing the fields $\delta_\eta(z)$ and $\delta_{\bar{\eta}}(\bar{z})$ does enlarge 
the algebra by the zero modes of $\xi$ and $\bar{\xi}$\ . Thereby, it gets extended to
to $\overline{N}(0,0)\oplus N(0,1^\star)\oplus N(1^\star,0)\oplus N^\star(1,1)\ $. 
This can be understood as follows:

In more detail,
$\overline{N}(0,0)=\l\bigoplus_{l\in\Z}\mc{A}_{\2}^-(l)\otimes
\overline{\mc{A}}^+_{-\2}(l)\r\otimes\overline{\bar{N}}(0)$
and the algebra of the auxiliary fermions does not factorize. 
Therefore, the logarithmic deformation must be performed on the factor
$\overline{\mc{A}}^+_{-\2}(l)\otimes\overline{\bar{\mc{A}}}^+_{-\2}(l')$ of each summand
and not on $\overline{\mc{A}}^+\otimes\overline{\bar{\mc{A}}}^+\ $, as before. In order to 
do so, we use (\ref{fieldtrafo}) which means that we act with 
$\delta_\eta(z)\ ,~\delta_{\bar{\eta}}(\bar{z})$ and 
$\delta_\eta(z)\delta_{\bar{\eta}}(\bar{z})$ on fields. From the point of view of the
Fock space, the algebra is thence shifted by $\xi_0\ ,~\bar{\xi}_0$ and 
$\xi_0\bar{\xi}_0$\ . Thereby we arrive at 
$\overline{\mc{A}}^+_{-\2}(l)\otimes\left.\bar{\mc{A}}^+_{-\2}\right.^\star(l'-1)$ via 
$\delta_\eta(z)$\ , and at
$\left.\mc{A}^+_{-\2}\right.^\star(l-1)\otimes\bar{\mc{A}}^+_{-\2}(l')$ and 
$\left.\mc{A}^+_{-\2}\right.^\star(l-1)\otimes\left.\bar{\mc{A}}^+_{-\2}\right.^\star(l'-1)$
by means of the other modes, respectively.

In terms of the full theory, the algebra just obtained is
$(\mc{F}_0\oplus\left.\mc{F}_0^1\right.^\star)\otimes(\mc{\bar{F}}_0
\oplus\left.\mc{\bar{F}}_0^1\right.^\star)\ $. 
As the CGO is defined on $\C_\times\ $, we apply a coordinate transformation to 
$\mc{F}^1_0\ $, ending up with the result just proposed. Notice, that 
$\mc{F}_0^1$ and $\mc{F}_\infty^1$ are glued together by means of $\mc{F}^1_\times$ and
therefore, $\mc{F}_\infty^1$ is naturally embeddeed in that fock space. 
 
We can now argue, why
the zero mode of the energy momentum tensor $T_\eta(z)$ of the full supersymmetric 
$bc$-system is a mapping 
$\left. T_\eta\right._0:~\mc{F}_\times^1\otimes\bar{\mc{F}}_\times^1\rightarrow
\l\mc{F}_\times^1\otimes\bar{\mc{F}}_\times^1\r
\oplus q\l\mc{F}_0\otimes\bar{\mc{F}}_0\r$\ .

In order to be more concrete, we define states
\begin{align}
  \begin{split}
    |n\>^{(l)}_0& :=\eta_{r_1}\cdots\eta_{r_i}\xi_{k_1}\cdots\xi_{k_j}|0\>_{\mc{A}^+}\ , ~~~
    \begin{array}{c}
      r_1<\cdots <r_i< 0,~~~k_1<\cdots <k_i< 0\ ,\\
      n=\sum |r_i|+|k_j|,~~~l=i-j
    \end{array}
  \end{split}
\end{align}
and
\begin{align}
  \begin{split}
    |n\>^{(l)}_\infty& :=
    \eta_{r_1}\cdots\eta_{r_i}\xi_{k_1}\cdots\xi_{k_j}|1\>_{\mc{A}^+}\ ,
    \begin{array}{c}
      r_1<\cdots <r_i< 0,~~~k_1<\cdots <k_i< 0\ ,\\
      n=\sum |r_i|+|k_j|,~~~l=i-j
    \end{array}
  \end{split}
\end{align}
These are basis elements of $\overline{\mc{A}}_{-\2}^+(l)$ and 
$\left.\mc{A}_{-\2}^+\right.^\star(l-1)$\ , respectively, for the zero modes of
$\eta$ and $\xi$ are absent. The algebra of the $|n\>^{(l)}_\infty$ has to be enlarged by 
the zero mode of $\eta\ $, such that $\left.\mc{A}_{-\2}^+\right.^\star(l-1)\rightarrow
\mc{A}_{-\2}^+(l-1)=\C[|n\>^{(l)}_\infty]_{n\in\N}\oplus
\eta_0\cdot\C[|n\>^{(l-1)}_\infty]_{n\in\N}\ $. Similar definitions hold for the 
anti-chiral states.

The modes of the deformed chiral energy momentum tensor of the auxiliary 
$\eta\xi$-system are $\left. T^{(a)}_\eta\right._k=T^{(a)}_k+q~\eta_k\bar{\eta}_0\ $. Their
action on $|n,\bar{n}\>^{(l,\bar{l})}_{s,\bar{s}}:=
|n\>_s^{(l)}\otimes|\bar{n}\>^{(\bar{l})}_{\bar{s}}\ ,~s,\bar{s}\in\{0,\infty\}$ 
is as follows. 

For the zero mode we have
\be
\left. T^{(a)}_{\eta}\right._0\cdot |n,\bar{n}\>^{(l,\bar{l})}_{s,\bar{s}}
= n|n,\bar{n}\>^{(l,\bar{l})}_{s,\bar{s}}-q\mc{N}\bar{\mc{N}}
|n,\bar{n}\>^{(l,\bar{l})}_{0,0}\ ,
\ee
where we introduce the shortcut
$\mc{N}:=(-)^{i+\bar{i}+j+\bar{j}}\delta_{s,\infty}$ and 
$\bar{\mc{N}}:=(-)^{i+\bar{i}+j+\bar{j}}\delta_{\bar{s},\infty}$\ . As $\eta_0^2=0$\ , 
the image of the Hamiltonian is in its kernel. Therefore, due to the first 
Grothendieck-Cousin operator,  
$\left. T_\eta\right._0:~\mc{F}_\times^1\otimes\bar{\mc{F}}_\times^1\rightarrow
\l\mc{F}_\times^1\otimes\bar{\mc{F}}_\times^1\r
\oplus q\l\mc{F}_0\otimes\bar{\mc{F}}_0\r$\ . 
On all other states in $\mc{F}_{\eta\bar{\eta}}$\ , it is diagonal.

For the other modes with $k\neq0$\ , we find
\be
\left. T^{(a)}_{\eta}\right._k\cdot |n,\bar{n}\>^{(l,\bar{l})}_{s,\bar{s}}
= T^{(a)}_k\cdot|n,\bar{n}\>^{(l,\bar{l})}_{s,\bar{s}}\pm q\eta_k\cdot\bar{\mc{N}}
|n,\bar{n}\>^{(l,\bar{l})}_{s,0} 
\ee
and the mapping $\left. T_\eta \right._k$ is in general not diagonal if the states are in
the subspace $\l\mc{F}_0\oplus\mc{F}_\times^1\r\otimes\bar{\mc{F}}_\times^1$\ . The
sign is plus if $s=0$ and minus, otherwise.

\subsection{Morse-Novikov Theory Included}
The results of the last section are unexpected from the point of view of the 
Morse-Novikov theory on $\widetilde{LX}$, that we are investigating. The reason is, 
that
there, we are only treating two sorts of spaces of states. One of them is associated with 
flow lines that descend from $\{0\}\in X$ and are 
described by polynomials of coordinates along $\widetilde{LX}_{0,n}$. The other
is related to flow lines descending from $\{\infty\}$, belonging to 
$\widetilde{LX}_{\infty,n}$. The Fock spaces on which they are modeled are, as we explained,
$\mc{F}_0\otimes\bar{\mc{F}}_0$ and $\mc{F}_\infty^1\otimes\bar{\mc{F}}_\infty^1$, 
respectively.

The space of states we obtained for the two dimensional field theory 
by means of the logarithmic deformation is, however, larger. It even includes 
states in $\l \mc{F}_0\otimes\bar{\mc{F}}_\times^1\r \oplus \l 
\mc{F}_\times^1\otimes\bar{\mc{F}}_0\r$\ that have no immediate geometric interpretation.
In a way, these spaces are mixing coordinate charts around $\{0\}\in X$ with charts around
$\{\infty\}\in X$, introducing an additional non-locality into the two dimensional field 
theory. 

\section{Discussion and Open Questions}

We will discuss our results in three graded steps. Let us start with a summary of the main 
results.

\subsubsection*{Summary}
We have deformed the supersymmetric $bc$-system with target and domain manifold 
being $\C\P^1\ $, associated to (\ref{gauged a model}). The guideline for 
our calculations was the condition that the zero mode of the energy momentum tensor should 
acquire additional terms
\begin{equation*}
T_0\stackrel{!}{\mapsto} T_0 + \delta_1 + \delta_2\ .
\end{equation*}
The argument has been, that the zero mode of the energy momentum tensor should be
identified with (part of) the Hamiltonian, the GCOs included, of the Morse-Novikov theory 
which is associated with the A-model as given by (\ref{gauged a model}). From the point of 
view of Morse-Novikov theory, these terms are an effect of the nontrivial geometry of 
target space $\widetilde{L\C\P^1}\ $.

The method of deformation was a logarithmic deformation of the supersymmetric
$bc$-system according to \cite{Fjelstad:2002ei}. We performed it for the first
GCO $\delta_1$ and started with the $bc$-system in the representation $\mc{F}_0\ $.
Thereby, the space of states was extended to 
$(\mc{F}_0\oplus\mc{F}_\times^1)\otimes(\bar{\mc{F}}_0\oplus\bar{\mc{F}}_\times^1)\ $.
The same applies to the second GCO and the full energy momentum tensor 
obtaines extra terms on every level:
\begin{equation*}
T_{-n}\mapsto T_{-n}+ \eta_{-n}\bar{\eta}_0 + q~
\tilde{\eta}_{-n}\bar{\tilde{\eta}}_0~,~~~n\geq 0\ .
\end{equation*}
The result is an LCFT with logarithmic 
partners on each level of the energy momentum tensor. The space of
states of the Morse-Novikov theory is included as a subspace, it is just 
$(\mc{F}_0\otimes\bar{\mc{F}_0})\oplus(\mc{F}_\infty^1\otimes\bar{\mc{F}}_\infty^1)\ $.
In terms of the original fields the energy momentum tensor reads
\begin{equation*}
T(z)\mapsto T(z) - \oint_0 \dd\bar{z}~\l \Psi_-(z,\bar{z}) +q\Psi_+(z,\bar{z})\r
\pi(z)\bar{\pi}(\bar{z})\ .
\end{equation*}
It is globally defined on $X$, where the chart transition has to come along with a change 
of sign in the $A$-field. 

The supercharge and hence the cohomology is also deformed and yields the globally
defined quantity:
\begin{equation*}
\tilde{Q}(z)\mapsto \tilde{Q}(z) + \frac{\i}{z} \oint_0 \dd\bar{z}~
\l \Psi_-(z,\bar{z}) - q~\Psi_+(z,\bar{z})\r \bar{\pi}(\bar{z})\ .
\end{equation*}

In short terms: we have shown on the level of the two dimensional supersymmetric 
$bc$-system, i.e.\ for the corresponding Virasoro algebra, that classically the proposal 
of Frenkel, Losev and Nekrasov is correct, the model satisfies all properties of an LCFT.
If we call it therefore an LCFT we mean that it has an appropriate Virasoro algebra.

\subsubsection*{Questions}
What are the questions that we could not answer? There are mainly two:

The more important one, which we did not touch, is if the thus deformed $bc$-system is an 
LCFT on quantum level? What is the vacuum expectation value $\<T_\eta\>$
of the deformed energy momentum
tensor and in which vacuum or in how many should we take it? Notice, that we considered the 
full moduli space of vacuum configurations of the theory with instantons and one of the
vacuum configurations is charged and not conformally invariant. As at least one of the 
GCOs is a mapping from the representation space of the 
vacuum which is not conformally invariant, we further are not sure in what respect we 
should call the quantum theory an (L)CFT.

The second, maybe related question results from our calculations. In order to make a
logarithmic deformation
we bosonized the theory. Thereby we introduced an auxiliary fermionic $bc$-system which
introduces an additional current and a charge anomaly. That is, however, invisible
in the perspective of the original theory. What effect does that have on the quantum level
and how is that related with the theory before bosonization?

\subsubsection*{Speculations}
So far the results and related questions. The motivation for the calculations we have done
was, however, the question 
we put forth in the introduction: ``can we understand the attribute of a conformal field 
theory on curved target space to be a {\em logarithmic} conformal field theory as an aspect 
of gravity?'' What can we say about that after all?

Of course we can not say anything about that on the level of the fully quantized
theory. On the classical level our r\'esum\'e is, that we can give a positive answer
because of the connection of non-unitarity, gravity and the presence of
only instantons that we have drawn. Especially for the logarithmically deformed $bc$-system 
it seems that its' non-unitarity gets a direct interpretation in terms of gravity, as we
explained in section \ref{the second}. Moreover, non-unitarity seemes to be a 
necessity for the supersymmetric $bc$-system to be an (L)CFT: on the level of the 
Virasoro algebra one can see
that a logarithmic deformation of the Hamiltonian of the $bc$-system can only be done by 
either $\eta_0$ or $\xi_0$ but not by both. This again mirrors that in the Morse-Novikov
theory under consideration only instantons (or if we choose, only anti-instanons) 
are present. 

The solitude of instantons results in a logarithmic deformation of the supersymmetric
$bc$-system by the GCOs whose presence can equivalently be interpreted as an effect of the 
non-trivial geometry of target space. 

\newpage
\section*{Acknowledgments}
Many people contributed to the work of KVs: First of all, I am indebted to Edward Frenkel.
Without his inspiring, critical comments and his integrity I could not have 
completed my study. Further I want to thank Matthias Blau for discussions and his 
amicable hospitality in Neuch\^atel. My colleagues and in particular Johannes Br\"odel, 
Andr\'e Fischer and Michael Klawunn helped me a lot with discussions and their 
encouragement. If I ``stood on the shoulders of giants'' it 
were mostly Barbara Duden's shoulders on which I tried to keep my balance.




\pagebreak
\appendix
\section{Notations and Conventions}
This appendix serves to fix the notations and to define the objects that are important
within this paper. We hope further, that it is self-contained enough in order to enable the 
ambitious reader, who may not be an expert on that subject, to understand the 
details left out in the main part and to reproduce the calculations.

\subsection{The CFT Conventions}\label{notation}
We will treat a special version of the topological $A$ model derived in 
\cite{Frenkel:2008vz}, consisting of a bosonic scalar field $x$, which is a mapping 
$x:~\Sigma\rightarrow X$ of a Riemannian surface into a CY manifold $X\ $. The embedding $x$ 
consequently 
splits into a holomorphic $x^a$ and an antiholomorphic part $x^{\bar{a}\ }\ $. Its dual $p$ 
splits into elements of $\Omega^{1,0}_\Sigma\otimes\Omega^{1,0}_X$ and 
$\Omega^{0,1}_\Sigma\otimes\Omega^{0,1}_X\ $, respectively: 
$p=p_{az}~\dd z\dd x^a+p_{\bar{a}\bar{z}}~\dd{\bar{z}}\dd x^{\bar{a}}\ $. 
However, we will not consider $p$ but 
$p':=p+\Gamma^a_b\psi^b\pi_a+\Gamma^{\bar{a}}_{\bar{b}}\psi^{\bar{b}}\pi_{\bar{a}}$, which
does not transform like a one form in $X$. The scalar $\psi$ and one form $\pi$ are the 
supersymmetric partner fields with the same transformation
properties as the bosons. Though they have the wrong statistics, we will call the 
superpartners ``fermions''.

\subsubsection{Basic Fields and the OPE Conventions}
The OPE structure for the embedding $x$ and its conjugate momentum is given as follows:
\begin{align}
  \begin{split}
    x^a(z)p'_b(\o)=\frac{\i\delta^a_b}{z-\o}\ ,~~~~
    \psi^a(z)\pi_b(\o)=\frac{-\i\delta^a_b}{z-\o}\ .
  \end{split}
\end{align}

\begin{center}
~$\star$~
\end{center}

We will also need a more general notation for OPEs. Therefore, let us introduce the
bracket $[\cdot,\cdot]_n$ according to \cite{Fjelstad:2002ei}. For any two conformal fields 
$A(z)$ and $B(\o)$ the element $[A,B]_n(\o)$ denotes a primary field in the OPE
\be
A(z)B(\o)=\sum_{n\geq 1}\frac{[A,B]_n(\o)}{(z-\o)^n}\ .
\ee
From that, we can derive the identity
\be
[B,A]_n(z)=(-)^{F_A F_B}\sum_{k\geq n}\frac{(-)^k}{(k-n)!}\partial_z^{k-n}[A,B]_k(z)\ ,
\ee
where $F_A$ denotes the fermion number of $A\ $.

\begin{center}
~$\star$~
\end{center}

The energy momentum tensor and current of interest will be
\begin{align}
  \begin{split}
    T(z)&=\i~:p'_a(z)\dz x^a(z)-\pi_a(z)\dz\psi^a(z):\ ,\\
    j(z)&=-\i~:x^a(z)p'_a(z)+\psi^a(z)\pi_a(z):\ .
  \end{split}
\end{align}
The conformal weights with respect to $T$ and charges are then
\begin{align}
  \begin{split}
    \Delta_T(x)&=0=\Delta_T(\psi)\ ,~~~\mbox{charge:}~-1\ ,\\
    \Delta_T(p')&=1=\Delta_T(\pi)\ ,~~~\mbox{charge:}~+1\ .
  \end{split}
\end{align}
Further we will need the supercharge, that is given as
\be
Q(z)=\i :p'(z)\psi(z):\ .
\ee

\subsubsection{Chiral Bosonization}\label{chiral bosonization}

Later, we will map that fields to a supersymmetric $bc$ system. Most of our conventions 
of that section goes back to \cite{Feigin:1990, Friedan:1985ge}. The $bc$ systems
corresponding to the bosons will be labeled by an index $\e=-$ and the one for the
fermions with an index $\e=+\ $. Its relations to the algebra above is given by
\begin{align}
  \begin{split}
    \e=-:&~~~x\mapsto b,~~~\i p'\mapsto c\\
     \e=+:&~~~\psi\mapsto b,~~~\i\pi\mapsto c\ .
  \end{split}
\end{align}

\begin{center}
~$\star$~
\end{center}

The $bc$-system above, gives rise to a Heisenberg algebra, generated by
\be
[c_n,b_m]=\delta_{n,-m}\ .
\ee
that has irreducible representations $M^\e(p),~p\in\Z\ $:
\be
b_n|p\rangle =0,~n>-p,~~~~c_n|p\rangle =0,~n\geq p\ .
\ee 
As above $M^\e(p)$ denotes the bosons by the choice of $+$ and otherwise the fermions. 
The current
\be
j^\e(z)=-:b(z)c(z):=\sum_{n\in\Z}j_n z^{-n-1}
\ee
gives a partition of $M^\e(p)=\bigoplus_{l\in\Z}M^\e(p)_l$ such 
that $|p\rangle\in M^\e(p)_{-\e p}\ $. The Virasoro algebra acts on $M^\e(p)$ by
\be
T^\e(z)=:\dz b(z)c(z):\ ,
\ee
where normal ordering is defined in the representation of $M^\e(0)\ $. 

Let  $u_n(z):=z^{-n+1}\partial_z$ and $\mc{L}_{u_n}$ denote the Lie derivative in
direction of $u_n$\ . The modes of the energy momentum tensor $T^\e_n$ act on the field 
modes $b_m$ and $c_m$ like the operator $T'_n=\mc{L}_{u_n}$ on the differentials
$z^{-m}$ and $z^{-m-1}\mr{d}z$\ :
\begin{align}
  \begin{split}
    [T_n^\e,c_m]&=-m~c_{m+n}\ ,\\
    [T_n^\e, b_m]&=-(m+n)~b_{n+m}\ .
  \end{split}
\end{align}
In particular, the Hamiltonian, i.e.\ the zero mode of the $bc$-system acts therefore like
the Lie derivative in direction of $z\partial_z$\ , 
the generator of the $\C^\times$-symmetry on $\C\P^1$\ , on each mode.

The central charge of $T^\e$
is given by $c^\e=-2\e\ $. Therefore, the central charge of the supersymmetric $bc$-system 
is zero. The current is anomalous with background charge $\mc{Q}=\e$:
\begin{align}
  \begin{split}
    T_n|p\rangle =0, n>0,&~~~~~T_0|p\rangle =\2\e~p(p-1)~|p\rangle\ ,\\
    j_n|p\rangle =0, n>0,&~~~~~j_0|p\rangle=-\e p~|p\rangle\ .
  \end{split}
\end{align}
In the case of fermions the extremal states are simply related by
\begin{align}
  \begin{split}
    |p+1\>&=\xi_{-p} \xi_{-p+1} \cdots\xi_0|0\>,~~~~~p\geq 0\ ,\\
    |p\>&=\eta_{-p}\eta_{-p+1}\cdots\eta_{-1}|0\>,~~~p<0
  \end{split}
\end{align}
and $\eta_0$ is mapping $\mc{M}^+(p)_l\rightarrow\mc{M}^+(p)_{l+1}\ $.

\begin{center}
~$\star$~
\end{center}

Chiral bosonization means that the algebra $M^\e(p)$ is expressed in terms of another
algebra $\mc{A}^\e_{\a_0}(h)\ $. This is the Heisenberg algebra
\be
[J_n,J_m]=\e~n\delta_{n,-m}
\ee
of a bosonic current $J^\e(z)=\sum_{n\in\Z} J_nz^{-n-1}$ with heighest wheight
representation given by
\be
J_n\nu_h=h\delta_{n,0}~\nu_h,~n\geq 0,~~~~h\in\C
\ee
and with an action of the Virasoro algebra by means of
\be
T_{J^\e}(z)=\e~\l \2:J^\e(z)^2:+\a_0\partial_zJ^\e(z)\r\ .
\ee
The central charge is $c_{J^\e}=\e(1-12\e\a_0^2)$ and the conformal wheight of $\nu_h$ is 
$\Delta_{T_{J^\e}}(\nu_h)=\2\e h(h-2\a_0)\ $. The identification is as follows. Set
$\a_0=-\2\e$ and introduce a chiral scalar field
\be
\phi^\e(z)
=\e\int^z J^\e(\o)\dd\o=\e~\l\phi_0+J_0\log{z}-\sum_{n\neq0}\frac{J_n}{n}z^{-n}\r,~~~
[\phi_0,J_n]=-\e \delta_{n,0}\ .
\ee
Define further vertex operators
\be\label{V}
V^\e(r,z)=:\exp{r\phi(z)}:=\mr{e}^{\e r\phi_0}z^{\e r\k}
\mr{e}^{-\e r\sum_{n<0}\frac{J_n}{n}z^{-n}}
\mr{e}^{-\e r\sum_{n>0}\frac{J_n}{n}z^{-n}}\ ,~~~~r\in\C\setminus\{0\}\ .
\ee
Then in the fermionic case
\begin{align}
  \begin{split}
    c(z)&\mapsto V^+(+,z)\ ,\\
    b(z)&\mapsto V^+(-,z)
   \end{split}
\end{align}
and $M^+(p)\simeq \mc{A}^+:=\bigoplus_{l\in\Z}\mc{A}_{-\2}^+(l)$ with 
$|p\>\mapsto \nu_{-p}\ $.
Notice that $M^+(p)\simeq M^+(p'),~\forall ~p,p'\ $. This is not true for the bosons and 
also the identification is not that simple.

In the bosonic case, the central charge does not fit with the one in the representation
$M^-(p)\ $. Therefore, one has to introduce an auxiliary fermionic $bc$-system, with 
$b:=\xi$ of conformal weight zero and $c:=\eta$ of conformal weight one and background 
charge
$\mc{Q}_{\eta\xi}=1\ $. The fields $b$ and $c$ of the former bosonic theory are then
identified by
\begin{align}
  \begin{split}
    c(z)&\mapsto V^-(+,z)\otimes V^+_{\eta\xi}(+,z)\ ,\\
    b(z)&\mapsto V^-(-,z)\otimes\partial_z V^+_{\eta\xi}(-,z)\ ,
  \end{split}
\end{align}
This corresponds, on the first sight, to the algebra
$N(p):=\bigoplus_{l\in\Z}~\mc{A}^-_{\2}(p+l)\otimes \mc{A}^+_{-\2}(l)\ $.
Since the zero mode of $V^+_{\eta\xi}(-,z)$ (or equivalently $\xi_0$) does not appear,
$M^-(p)$ is, however, only isomorphic to the kernel of the mapping 
$1\otimes V^+_{\eta\xi,0} :N(p)\rightarrow N(p-1)$ (or equivalently the kernel
of $1\otimes\eta_0\ $). Therefore, $M^-(p)\simeq
\overline{N}(p):=\bigoplus_{l\in\Z}~\mc{A}^-_{\2}(p+l)\otimes
\overline{\mc{A}}^+_{-\2}(l)\ $,
where the overline denotes the kernel of $\eta_0\ $. This is much nicer and in more detail 
described in \cite{Feigin:1990}.

\subsection{The Chiral De Rham Complex}

The chiral de Rham complex is introduced in  \cite{Schechtman:1999}. It is a sheaf of
vertex algebras on a smooth manifold $X$ with a canonical embedding of the de Rham complex.
The starting point are the Heisenberg and Clifford algebras of the supersymmetric 
$bc$-system. The zero modes $\{x^a_0, \psi^a_0\}$ and $\{\left.p'_a\right._0, 
\left.\pi_a\right._0\}$ can be identified with geometric objects on $X$ in the following 
manner:
\begin{equation*}
\begin{tabular}{ll|ll}
bosons:&~&fermions:&~\\\hline
$x^a_0$& $x^a$&$\psi^a_0$&$\dd x^a$\\
$\left.p'_a\right._0$& $\partial_a$&$\left.c_a\right._0$&$\iota_a$
\end{tabular}
\end{equation*}
In that respect, we can define an exterior derivative, which is just the
de Rham differential by $\dd:= Q_0\ $, where $Q_0$ is the supercharge. Chart transitions
can be defined on the zero modes and that supplies us with the de Rham complex. The
chiral de Rham complex generalizes these chart transitions to all modes and to the
fields that can be assigned to them. They read: 
\begin{align}\label{trafo rules}
  \begin{split}
    x\mapsto\phi_x(z)&=g(x)(z)\ ,\\
    p'\mapsto\phi_{p'}(z)&=\l \frac{\partial g^{-1}}{\partial\phi_x}p'
    +\frac{\partial^2 g^{-1}}{\partial\phi_x^2}\frac{\partial g}{\partial x}
    \pi\psi\r (z)\ ,\\
    \psi\mapsto\phi_\psi(z)&=\frac{\partial g}{\partial x}\psi(z)\ ,\\
    \pi\mapsto\phi_\pi(z)&=\frac{\partial g^{-1}}{\partial \phi_x}\pi(z)\ .
  \end{split}
\end{align}
Especially one can show that, when $\phi_x=x^{-1}\ $, it is only the zero mode $x_0$
that has to be inverted \cite[pg.~91]{Frenkel:2008vz}. The composite fields, like the 
energy momentum tensor and the supercharge undergo
transformations
\begin{align}
  \begin{split}
    T(z)&\mapsto\phi_T(z)=T(z)\ ,\\
    Q(z)&\mapsto\phi_Q(z)=Q(z)-\partial_z\l \frac{\partial}{\partial{\phi_x}}
    \left[\mr{log}~\frac{\partial{g^{-1}}}{\partial{\phi_x}}\right]\phi_\psi(z)\r\ .
  \end{split}
\end{align}

\subsubsection{The Fields $\eta$ and $\xi$ in Terms of the Original 
Fields}\label{in original}
The fields $\eta$ and $\xi$ can be expressed in terms of the original fields.
That has already been calculated in \cite{Frenkel:2008vz} for $\eta$. Below, we will 
summarize how that works and also give the precise formulas for $\xi$ and
the supercharge.
 
The trick is the following. Start with the original fields in the coordinate chart
$\C_0\ $. Apply a coordinate transformation on $X$: $~g: X\ni x\mapsto g(x)=\mr{exp}(x)\ $. 
All fields transform according to 
(\ref{trafo rules}):
\begin{equation}\label{field trafo}
  \begin{array}{ll}
    \phi_x(z)=\mr{e}^{x(z)}\ ,&\phi_{p'}(z)=\mr{e}^{-x(z)}\l p'(z)+\psi(z)\pi(z)\r\ ,\\
    \phi_\psi(z)=\mr{e}^{x(z)}\psi(z)\ ,& \phi_\pi(z)=\mr{e}^{-x(z)}\pi(z)\ ,\\
    \phi_T(z)=T(z)\ ,& \phi_Q(z)=Q(z)+\partial_z\psi(z)\ .\\
  \end{array}
\end{equation}
Exponentiating $x$ generalizes its definition to
$x(z)\mapsto x(z)+\hat{\o}~\mr{log}~z$\ . This demands an additional zero mode
conjugate to $\hat{\o}$\ , that we get by substituting $p'(z)=\partial_z U(z)\ ,~
U(z)=\int^z p'(\o)\dd\o$\ .\footnote{This leads to a different representation space, as 
described for example in \cite{Frenkel:2005ku} and \cite{Borisov:1998}. For our task, that
is, however, not of importance.} Now, we can bosonize the model in analogy with
\cite{Frenkel:2008vz} and thereby obtain
\begin{align}
  \begin{split}
    \eta(z)&\mapsto \i\pi(z) \mr{e}^{-\i\int^z p'(\o)\dd\o}\ ,\\
    \xi(z)&\mapsto \psi(z) \mr{e}^{\i\int^z p'(\o)\dd\o}\ .
  \end{split}
\end{align}
The Grothendieck-Cousin operators are obtained after a change to $\C_\times\ $, for they
are mappings between states in the chart $\C_0$ and $\C_\infty\ $, respectively. 
The corresponding transformation $\phi_x\mapsto \phi_x^{-1}$ has the effect that each 
element 
of $\{p',\psi,\pi\}$ gets multiplied by by $-1$\ . The Grothendieck-Cousin operators we 
proposed are now written as
\begin{align}\label{GC op}
  \begin{split}
    \delta_{(\infty,0),(0,1)}&=-q\oint_{0,0} \Psi_+(z,\bar{z})\pi(z)\bar{\pi}(\bar{z})\ ,\\
    \delta_{(0,0),(\infty,0)}&=\oint_{0,0} \Psi_-(z,\bar{z})\psi(z)\bar{\psi}(\bar{z})\ ,
  \end{split}
\end{align}
where
\be
\Psi_\pm(z,\bar{z})=\Psi_\pm(z)\Psi_\pm(\bar{z})=\mr{e}^{\pm\i\l\int^z p'(\o)\dd\o+
\int^{\bar{z}}\bar{p}'(\bar{\o})\dd\bar{\o}\r }\ .
\ee
The supercharge transforms to
\be
\phi_Q(z)\mapsto \tilde{Q}(z)=Q(z)-\partial_z\psi(z)\ ,
\ee
while the energy momentum tensor is again not affected.

\subsection{Morse-Bott-Novikov Theory on Loop Space}\label{loop space}
A nice source on that topic is \cite{Floer:1989}. The case we will need is that $X$ is a 
simply connected, compact K\"ahler manifold.
The loop space $LX$ of $X$ is the space of smooth maps $\gamma:~S^1\rightarrow X\ $. 
It is not simply connected. If we want to formulate a Morse theory on loop space, 
that is a problem because the potential $v(x)$ is supposed to be a gradient of some 
function $f\in C^\infty(X)\ $. Therefore, on lifts the model to the universal cover, which 
is the space
\be
\widetilde{LX}:=\left\{(\g,\tilde{\g})\left.\right|~\g\in LX,~\tilde{\g}:D\rightarrow X~
\mbox{such that}~\left.\g=\tilde{\g}\right|_{\partial D}\right\}/ \sim\ ,
\ee
where $\sim$ denotes equivalence with respect to homotopy. On $\widetilde{LX}\ $, $f$ can
be defined uniquely according to
\be
f(\tilde{\g})=\int_D\tilde{\g}^*(\omega_K)-\a\int_{\partial D}\g^*(H)\ .
\ee
$\omega_K$ denotes the K\"ahler two-form and $H_\a$ is a smooth, real valued function on
$\partial D\times X\ $, reducing the critical manifold of the Hamiltonian flow equation to
isolated points on $X\ $. If $\a=0\ $, we have the situation of a Morse-Bott-Novikov
theory, otherwise, it is Morse-Novikov. On $LX\ $, $f$ is a multivalued function and the
sheets of the covering, i.e.\ $\widetilde{LX}_n\ $, are counted by $H_2(X,\Z)\ $. That 
appears within the space of states
as an equivariance operator $q^n~,~n\in\Z\ $, \cite{Frenkel:2008vz}.

Let $\a\neq 0$ and $x_\mu$ be a critical point of the flow equation $\phi_v(x,t)=x(t)\ $. 
The descending manifolds $\widetilde{LX}_{\mu,n}$ are defined as
\be
\widetilde{LX}_{\mu,n}:=\left\{\tilde{\g}\in\widetilde{LX}_n~:~\lim_{t\rightarrow -\infty}
\phi_v(\tilde{\g},t)=x_\mu\right\}\ .
\ee

\end{document}